# Catalyst-Free And Morphology-Controlled Growth Of 2D Perovskite Nanowires for Polarized Light Detection

*Debjit Ghoshal[1], Tianmeng Wang[1], Hsin-Zon Tsai[2], Shao-Wen Chang[2], Michael Crommie[2], Nikhil Koratkar[3, 4]\* and Su-Fei Shi[1, 5]\**

[1]Department of Chemical and Biological Engineering, Rensselaer Polytechnic Institute, Troy, NY, 12180, United States.

[2]Department of Physics, University of California at Berkeley, Berkeley, CA, 94720, United States.

[3]Department of Mechanical, Aerospace and Nuclear Engineering, Rensselaer Polytechnic Institute, Troy, NY, 12180, United States.

[4]Department of Material Science and Engineering, Rensselaer Polytechnic Institute, Troy, NY, 12180, United States.

[5]Department of Electrical, Computer, and Systems Engineering, Rensselaer Polytechnic Institute, Troy, NY, 12180, United States.

\*shis2@rpi.edu, koratn@rpi.edu.

**Two-dimensional (2D) perovskites or Ruddleson Popper (RP) perovskites have emerged as a class of material inheriting the superior optoelectronic properties of two different materials: perovskites and 2D materials. The large exciton binding energy and natural quantum well structure of 2D perovskite not only make these materials ideal platforms to study light-matter interactions, but also render them suitable for fabrication of various functional optoelectronic devices. Nanoscale structuring and morphology**

**control have led to semiconductors with enhanced functionalities. For example, nanowires of semiconducting materials have been extensively used for important applications like lasing and sensing. Catalyst-assisted Vapor Liquid Solid (VLS) techniques, and template assisted growth, have conventionally been used for nanowire growth. However, catalyst and template-free scalable growth with morphology control of 2D perovskites have remained elusive. In this manuscript, we demonstrate a facile approach for morphology-controlled growth of high-quality nanowires of 2D perovskite, $(BA)_2PbI_4$. We demonstrate that the photoluminescence (PL) from the nanowires are highly polarized with a polarization ratio as large as ~ 0.73, which is one of the largest reported for perovskites. We further show that the photocurrent from the device based on the nanowire/graphene heterostructure is also sensitive to the polarization of the incident light with the photocurrent anisotropy ratio of ~3.62 (much larger than the previously reported best value of 2.68 for perovskite nanowires), thus demonstrating the potential of these nanowires as highly efficient photodetectors of polarized light.**

Organic Inorganic Halide Perovskites (OIHP) or hybrid perovskites have gained immense attention in the last few years due to its excellent electronic and optical properties.[1–4] These materials have found applications in low-cost, highly efficient energy harvesting devices.[5–8] They have also shown exceptional performance in optoelectronic devices like light emitting diodes (LEDs) and detectors.[9–12] However, these materials suffer from their lack of stability in ambient conditions (oxygen and moisture) and their performance degrades with time.[13–16] Two dimensional derivatives of perovskites, with hydrophobic organic spacers separating the lead halide octahedrons, have led to the emergence of a new class of materials with enhanced stability and functionality.[17–20] These 2D perovskites, in addition to being more stable, also show high quantum yields and high binding energy of excitons, hence serving as ideal platforms to study and manipulate light-matter interactions at the nanoscale.[21–23] 2D perovskites, with their natural quantum confined layered structure, ideally bridge the gap between these two emerging classes of materials with exceptional optoelectronic properties: 2D materials and perovskites, bringing in the best of both worlds with superior in-plane mobility, high binding energy and enhanced stability.[24]

Morphology control has proven to be an effective strategy in realization of semiconductors with greater potential.[25] Nanoplatelets of semiconductors have been used as high gain media and observation of whispering gallery modes.[26–28] Nanowires of different materials have been used in various applications like photodetectors, sensors and field emitters.[29–31] As such, morphology controlled growth of 2D perovskites is crucial for enabling better use of the material in optoelectronics. Additionally, nanowires of 3D perovskites have demonstrated low threshold lasing due to dielectric confinements of the excitons in such structures.[29,32] Such confinement effects are expected to be much stronger in 1D nanowire architecture of 2D perovskites due to their layered structure leading to more pronounced confinement effects. The ability to reproducibly synthesize nanowires could facilitate coupling the gain media and the cavity within the same material leading to more compact systems for lasing. However, template-free and scalable growth of nanowires of 2D perovskites has not been realized due to lack of understanding in the fundamental growth mechanism for these materials and lack of techniques that enable morphology controlled growth.

Solution based techniques have been conventionally used for the growth of 2D perovskites. Although solution based techniques offer a cheap route for the manufacturing of these layered perovskites, they limit the integration of these materials into the semiconductor industry where vapor based vacuum processing is preferred. These solution based processes also suffer from lack of controllability and reliability. Low Temperature Vacuum Assisted Chemical Vapor Deposition (LTCVD) based on "powder vaporization" methods represents the most scalable approach for growth of semiconductors and has extensively used for scalable synthesis of 2D transitional metal dichalcogenides (TMDs). This technique allows more controllable growth of 2D semiconductors by tuning various knobs for growth hence ensuring better control over thickness, morphology and composition of the material grown. This catalyst free growth of nanowires leads to nanowires with better properties and dimensionality as opposed to catalyst assisted traditional VLS techniques.

In this paper, we demonstrate a facile strategy for the morphology control of the 2D perovskite Butylammonium Lead Iodide ($(BA)_2PbI_4$) by changing the concentration of the precursors through simply changing the distance of the substrate from the source precursors by LTCVD. This paves the way for the controllable catalyst and template free growth of 2D $(BA)_2PbI_4$ nanowires of high quality, which can be potentially exploited for lasing applications. Additionally, we show that the optical response of the nanowires is highly sensitive to the angle between the light polarization and nanowire orientation. Finally, we exploit the polarization sensitive absorption of these nanowires to demonstrate the application of these as highly sensitive polarization photodetectors.

The 2D $(C_4H_9NH_3)_2PbI_4$ in our study is grown by our home-built LTCVD setup. A brief description of the growth process is as follows. The CVD growth substrate is placed downstream on the quartz tubes. Precursors Lead Iodide ($PbI_2$) and n-Butylammonium Iodide ($CH_3NH_3I$) are heated in a ~1 inch tube furnace at a temperature of ~400$^0$C. The furnace is purged with ultrahigh purity (UHP) Argon for ~10 minutes prior to the reaction and UHP Argon maintained at ~30 sccm is used as the transporting gas during the reaction. The temperature ramping rate is ~30$^0$C/min and the growth temperature is maintained for ~10 minutes. The pressure inside the furnace is maintained at ~600 mtorr during the growth process. A schematic of the growth process as well as the temperature ramping rates are illustrated in **Figure S1** and **Figure S2**. **Figure 1a** shows the different regimes of crystal growth. While supersaturation is an essential condition for crystal growth, the degree of supersaturation plays a crucial role in dictating the morphology of the crystal. **Figure 1b** shows a schematic of the growth process demonstrating a gradient in concentration across the length of the furnace allowing for the growth of samples of different morphology as shown in **Figure 1 c-e**. We show that while substrates closest to the precursors grow as nanoplatelets (Figure. 1c), the ones farthest away grow as nanowires (Figure 1e). Figure 1d shows the morphology at intermediate positions. The absence of spherical tips at the edge of the nanowires rule out VLS as the mechanism for growth. High resolution scanning electron microscopy (SEM) imaging (**Figure S3**) with of some of the nanowire samples show spiraling nature of the growth in some of these nanowires suggesting screw dislocation as a likely mechanism for nanowire growth however further experiments need to be done for confirmation of the mechanism.At very high initial concentration of precursors, dendritic growth was also observed at regions of highest concentrations similar to the findings of Song et al.  (**Figure S4**).[33]In stark contrast to the conventional catalyst assisted VLS technique for nanowire growth, where diameter of the nanowire is dictated by the dimensions of the metal catalyst, our catalyst independent method offers better control over the dimensions of the nanowire. **Figure S5** shows the SEM images for distribution of nanowires of different dimensions grown by LTCVD. **Figure S6** shows higher magnification SEM images showing the large difference in diameters of nanowires grown by LTCVD demonstrating good control of width of the nanowire using this technique. Additionally the catalyst-free growth ensures cleaner nanowires with less contamination resulting in enhanced optoelectronic properties.

We further perform Atomic Force Microscopy (AFM), Scanning Electron Microscopy (SEM) and Energy Dispersive Spectroscopy (EDS) to investigate the quality of the nanowires by looking into their morphology and stoichiometry. AFM measurement of one such nanowire shows the step height shown in **Figure 2a** which reveals the thickness of the nanowire to be ~26.6 nm. The 3D AFM image of the same nanowire is shown in **Figure S7**. The AFM topography image and 3D image demonstrate uniform thickness across the length of the nanowire. Field Emission Scanning Electron Microscope (FESEM) images of an individual nanowire are shown in **Figure 2b**. The clean facets of the nanowire point to single crystalline nature of the nanowire. EDS provides quantitative information about the elemental composition of the nanowire, and our measurement reveals the stoichiometry of the sample with Pb:I ratio approximately 1:4, in excellent agreement with our expectation (**Figure S8**). EDS spatial mapping (Figure. 2c-e) shows the uniform distribution of C, I and Pb across the dimensions of the nanowire demonstrating uniformity in chemical composition across the length of the sample. Similar results were found for the nanoplatelets as well which showed a uniform distribution of the constituent elements across the nanoplatelets (**Figure S9**).

$(C_4H_9NH_3)_2PbI_4$ has organic spacers separating the inorganic lead halide layers, as schematically shown in **Figure S10**. This gives it a natural quantum well structure resulting in enhanced optical properties (**Figure 3a**).[34] We next investigate the optical properties of the as-synthesized nanowires. **Figure 3b** shows the power dependence of the room temperature PL of the nanowires. The PL spectra of the nanoplatelets exhibit similar behavior to the nanowires at room temperature (**Figure S 11a**), with peaks centered at 519 nm. The inset shows that the power dependence has two different regimes following different power laws below and above a threshold excitation power. The integrated PL intensity is proportional to the power of the incident laser ($I \alpha P^n$) but with different values of n below and above a threshold power. Fitting in the lower excitation power regime shows a linear trend while fitting in the higher excitation power regime follows sub-linear (n=0.3) power law demonstrating saturation behavior. The sub linear power law suggests absorption saturation in the material beyond the threshold power. **Figure 3c** shows the power dependence at low temperatures (77 K). The power dependence is observed to be linear when excited with excitation power below 20 µW. At low temperatures, higher excitation powers were not used due to the possibility of sample damage due to reduced thermal conductivity at low temperatures leading to enhanced sample heating. The inset shows the bright florescence image of a nanowire pointing to the high quality of the samples. The PL peaks are much sharper at 77 K with a full width half maximum (FWHM) of 5 nm. The low-temperature PL spectrum reveals different behaviors for the nanowires compared to the nanoplatelets (**Figure S 11b**). While we see two peaks at 485 nm and 515 nm for the nanoplatelets, only one peak at 518 nm is observed for the nanowires. The excitonic peak shifts from 515 nm in nanoplatelets to 518 nm in nanowires can be attributed to the higher excitonic binding energy in the nanowires due to quantum confinement and decreasing of dielectric constant. This is consistent with the previous reports which suggested thin nanoflakes of exfoliated samples from single crystals also show two peaks. The origin of the two peaks has been attributed to the existence of 2 phases in the perovskite.[23] However, the presence of only one peak in the nanowires at 77 K indicates that only one phase is stabilized in the nanowires as opposed to nanoplatelets, and thin flakes exfoliated from bulk single crystals thus ensures superior performance of the nanowires in optoelectronic devices. Time-resolved photoluminescence (TRPL) was done to get dynamic information about the excited photocarriers. **Figure 3d** shows the decay in PL intensity measured using an Avalanche Photodetector (APD) when excited with a pulsed laser with a repetition rate of 40 MHz centered at 430 nm. A single exponential fit was used to estimate the lifetime which was found to be approximately 0.5 ns, which is longer than the typical lifetime of 250 ps reported for $(BA)_2PbI_4$.[35] The long lifetime compared to previous literature demonstrates high quality of the sample for optoelectronic applications.

Polarization dependent PL measurements shows that the absorption, and hence emission, is highly dependent on the polarization of the incident light, with respect to the orientation of the nanowire. For extremely narrow nanowires (nanowire width less than Bohr radius of exciton), the origin of anisotropy in PL emission has been attributed to quantum effects. However, since Coulomb interaction in 2D perovskites much stronger (indicated by the large binding energy), we expect the Bohr radius of the exciton in 2D perovskite to be much smaller compared to the diameters of the thinnest nanowires we synthesized (about 50 nm)[3]. Hence we attribute the origin of anisotropic response in our narrow nanowires to dielectric confinement effects. **Figure 4a** shows that the emission is the strongest when the nanowire is excited with a linearly polarized light parallel to the axis of the nanowire and weakest when it is perpendicular to the axis of the nanowire. Red curve in Figure 4a shows that the cosine fitting of the emission spectrum with a period of 2π. **Figure 4b** shows a color plot demonstrating the variation in the PL spectra of the nanowire as a function of the polarization angle. The single sharp peak at around 518 nm shows maximum intensity at an angle of $150^0$ (parallel to the orientation of the nanowire) and lowest intensity at $60^0$ and $240^0$ (perpendicular to the nanowire) demonstrating the correlation between the orientation the nanowire with the polarization of the incident light and the emission intensity. This large polarization anisotropy in emission from nanowires with angle of incident linearly polarized light with respect to the nanowire is consistent with the model of nanowires from previous literature.[36]. **Figure 4c** shows the PL emission spectrum along parallel and perpendicular directions. The emission polarization ratio, defined as $(I_{max}-I_{min})/(I_{max}+I_{min})$, is found to be as high as 0.73 for extremely narrow nanowires where $I_{max}$ and $I_{min}$ are the maximum and minimum values of PL intensities at the corresponding polarization angles. . In order to investigate the effects of nanowire width on polarization dependence of PL,we performed measurements on nanowires with different diameters. Polarization-dependent PL measurements were performed on wider samples of $(BA)_2PbI_4$. While the wider samples showed the same PL peak position, it was found that the polarization ratio decreased. Extremely wide samples (approximately 2-3 μm in diameter) (Figure S12 a) showed low values of polarization ratio (approximately 0.15) due to negligible effects of dielectric confinements, as the diameter of the nanowire was larger than the wavelength of excitation laser. Samples with smaller diameter (approximately hundreds of nanometers in diameter) showed higher polarization ratio of about 0.42 and 0.73, as shown in Figure S12 b and Figure 4 a, consistent with our interpretation.

The large optical anisotropy in emission opens up applications of the material in all optical switches, logic gates and various other applications. Here we demonstrate polarized photo-detection with these RP perovskites (n=1) by building a graphene/ $(BA)_2PbI_4$ heterostructure in a graphene field effect transistor (FET) device configuration. The architecture of the device is schematically shown in **Figure 5a**. We first transfer CVD grown graphene on copper film to the Si substrate with 285 nm thermal oxide, and we then grow 2D perovskite nanowires on graphene. Finally, gold electrodes are deposited on graphene as contact electrodes using a shadow mask. In the perovskite/graphene hybrid structure, the perovskite acts as a photoactive light absorbing layer and the graphene works as a current transport layer. The photoexcitation of the perovskite alter the conductivity of graphene underneath and hence induce the photoconductivity change of the graphene. The induced photocurrent was measured by a lock-in amplifier which is synchronized to a mechanical chopper which modulates the light intensity. The details of the experimental setup are described in the methods section. **Figure 5b** shows the photocurrent from the device as a function of the source drain voltage. The photocurrent is found to be linearly proportional to the source drain voltage, consistent with the optical excitation induced conductivity change of graphene. To further confirm that the photocurrent response stems from the optical absorption of the perovskite, we measure the photocurrent response as a function of the excitation wavelength, as shown in **Figure 5c**. The small bump at 518 nm is consistent with the PL peak of the perovskite, which corresponds to the exciton absorption of the perovskite and confirms the photocurrent response from the hybrid device originates from the optical response of the perovskite. Interestingly, the photocurrent response increases at lower wavelength and is peaked at ~ 480 nm. The pronounced photocurrent peaks suggests a strong absorption of light by perovskite at ~ 480 nm. This large absorption might due to trap states which are long lived due to localized electrostatic screening provided by the BA cations and $PbI_4$ anions.[37] The large absorption of the perovskite in the UV regime, combined with the high mobility of graphene renders our hybrid device a promising candidate for efficient UV detectors.

We also modulate the hybrid device with on-resonance (photoexcitation at 492 nm) and off-resonant (photoexcitation at 570 nm), and the results are shown in **Figure 5d**. While the on-resonance excitation generates large photocurrents (red curve in Figure 5d), the off-resonance excitations induce a smaller response (black curve in Figure 5d). It is worth noting that the photocurrent response rises quickly after the optical excitation, with the rise time <2 secs, limited by the time constant we used for the lock-in measurement. Finally we demonstrate the application of the hybrid device (**Figure 5e**) as a polarized light detector. Upon the excitation of a linearly polarized light pulse laser with optical excitation centered at 2.52 eV, the photocurrent of the hybrid device is a strong function of the angle between the polarization of the incident light and orientation of the nanowire. Similar to the PL spectra, the photocurrent is maximum when the polarization of light is parallel to the nanowire and minimum when it is perpendicular to it. The anisotropy ratio which is defined by $I_{max}/I_{min}$ is found to be 3.62 much larger than the previously reported best value 2.68 amongst the family of perovskite nanowires.[38]

In conclusion, we for the first time demonstrate a template-free controlled growth of 2D perovskites nanowires by LTCVD method. We demonstrate high-quality nanowires of uniform thickness and composition, along with exceptional optical properties. We also reveal that the emission from the nanowires is highly dependent on the incoming polarization of the incident light owing to anisotropic absorption of the nanowires. The polarization ratio is found to be as high as 0.73, one of the largest reported for perovskites.[39] Finally, we fabricate a hybrid $(C_4H_9NH_3)_2PbI_4$ /graphene/hybrid device and demonstrate its application as a polarization photodetector with photocurrent anisotropy ratio of 3.62.

**Experimental Section**

*Synthesis of Nanowires and Nanoplatelets Of $(C_4H_9NH_3)_2PbI_4$:* $(C_4H_9NH_3)_2PbI_4$ is grown in a CVD reactor (Figure 1a) as described in the main manuscript with and $PbI_2$ precursors maintained at separate temperatures. The growth temperature profile is provided in Figure S2. $SiO_2$ was used as a substrate for CVD growth. Substrates were cleaned with Piranha solution prior to growth.

*Graphene Transfer And Growth On Graphene:* Graphene grown on Copper was transferred onto 285 nm SiO2/ Si substrate using standard wet transfer techniques. Poly(methyl methacrylate) (PMMA) A4 is spin coated onto graphene on Copper substrate at 1500 rpm for 60 seconds followed by baking at $120^0$C after each step of spin coating. Copper is etched from the spin coated material using ammonium persulfate solution (3g in 50 ml of water). The graphene/PMMA stack is scooped from the solution after the copper is completely etched using a Si substrate followed by a series of rinsing in DI water. Finally the rinsed stack is scooped from DI water using 285 nm Si/$SiO_2$ substrate. The stack is dried in air for a few minutes. In the final step, the PMMA is dissolved in acetone and the graphene/substrate stack is rinsed in Isopropanol and dried with Nitrogen. This perovskite nanowires are then grown on graphene as described in the main text.

*Materials Characterization:* Scanning Electron Microscopy (SEM) and Electron Dispersive Energy Spectroscopy (EDS) was performed by using a ZEISS SUPRA 55 Field emission scanning electron microscope. Tapping mode of Multimode AFM from Digital Instruments was used for obtaining the topography images.

*Optical and PL Measurements:* Photoluminescence (PL) was measured with a home-built confocal microscope setup using lasers under different excitation conditions depending on the type of measurement. The details of the type of laser used, wavelength and excitation power have been listed in the main manuscript. A spectrograph (Andor) and a thermoelectric cooled CCD camera (Andor) was used for the spectroscopy measurement. The time-resolved PL (TRPL) was measured through the time-correlated single photon counting (TCSPC) technique and an avalanche photodiode detector (APD, by Micro Photon Devices) was used. A pulsed laser centered at 430 nm was used as the excitation source. For both PL and TRPL measurements, the excitation lasers were focused to a spot size with the diameter of 2 μm. Continuous wave (CW) diode laser (405 nm) was also used as an excitation source for steady state PL measurements

.

*Polarization Dependent PL Measurements:* Polarization Dependent measurements were carried by exciting the sample with linearly polarized light. A polarizer was used in the incident light path to excite the sample with linearly polarized light. The emitted light from the sample was again passed through a polarizer. The angle of the second polarizer was changed to record the intensity of the collected light along different polarizations.

*Polarization Dependent Photocurrent Measurements:* Photocurrent measurements were done using a lock-in amplifier to obtain better signal to noise ratios. A chopper was used to modulate the incident laser on the sample and this was used as a reference signal for the current amplifier. For wavelength dependent measurements, where the power of the incident laser varied due to change in wavelength, all the signals were normalized using a standard Si photodetector. Polarization dependent measurements were done as described in the previous section and are shown in **Figure S13**.

**Acknowledgements**
S.-F. S. and N.K. acknowledge funding support from the US National Science Foundation (Award 1608171). M.C would like to acknowledge funding support from the US National Science Foundation DMR-1807233. D.G. acknowledges the Center for Materials, Devices and Integrated Systems and the Center for Future Energy Systems at Rensselaer Polytechnic Institute for access to their facilities.

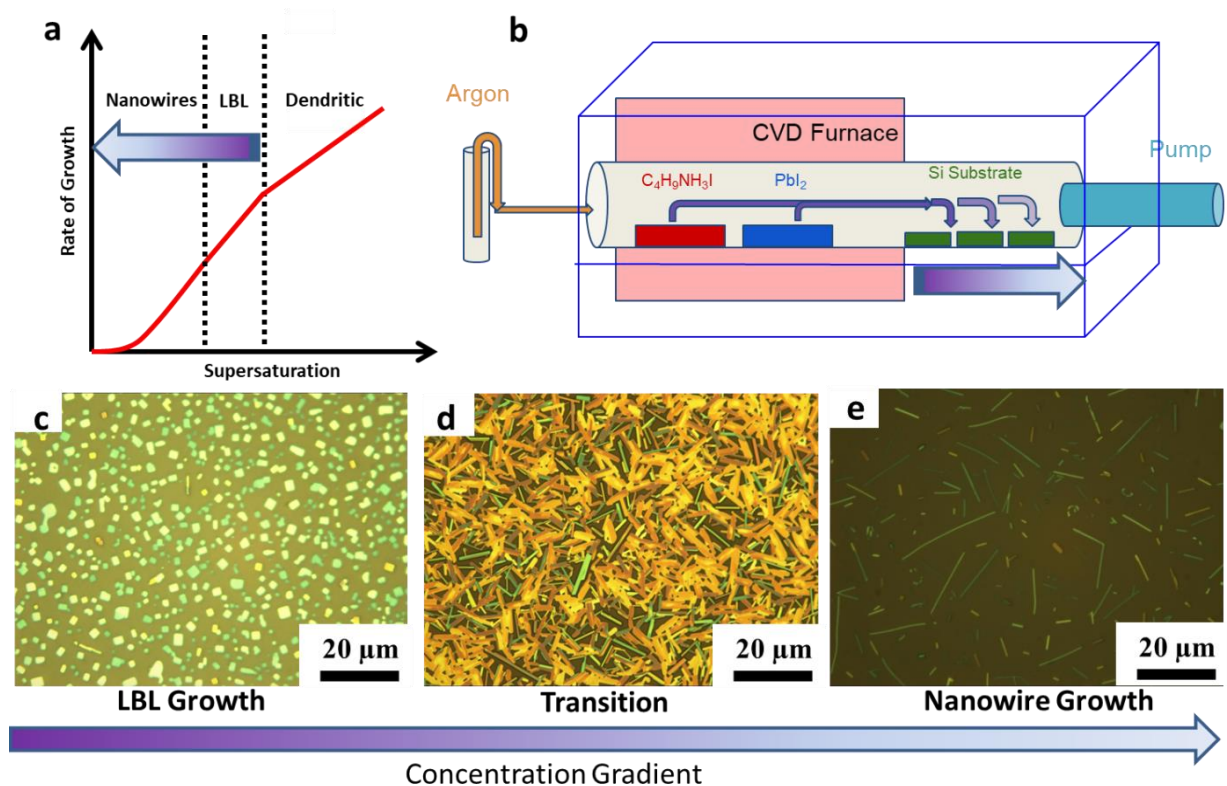

**Figure 1.** 2D perovskite morphology evolution with position. (a) Regimes of growth as a function of supersaturation, (b) Schematic of the growth process, (c-e) Optical microscope images of shape evolution from Nanoplatelets to Nanowires. Part (c) shows Layer by Layer (LBL) Growth at high precursor concentrations which transforms to nanowire growth at lower precursor concentrations, as shown in part (e).

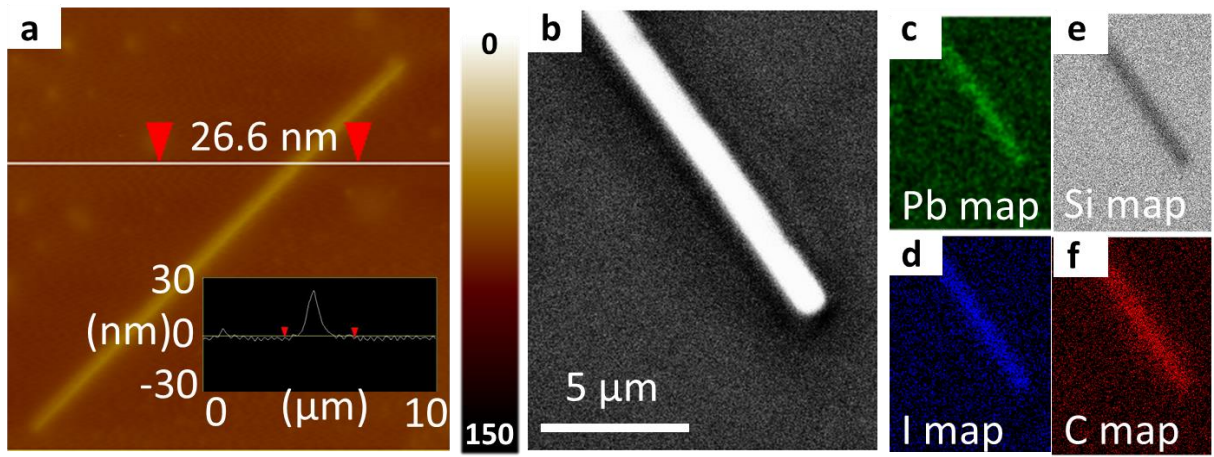

**Figure 2.** Characterization of 2D perovskite nanowires. (a) Atomic Force Microscope topography measurement shows uniform thickness of the nanowire. Inset: Line scan across the nanowire showing the thickness of the nanowire to be ~26.6 nm. (b) Scanning Electron Microscope image of a nanowire. (c-f) Elemental maps of Lead, Iodine, Silicon and Carbon showing uniform distribution of the elements Lead, Iodine and Carbon across the length of the nanowire. The Silicon elemental map shows reduced signal across the length of the nanowire due to reduced signal from the substrate due to the presence of the nanowire.

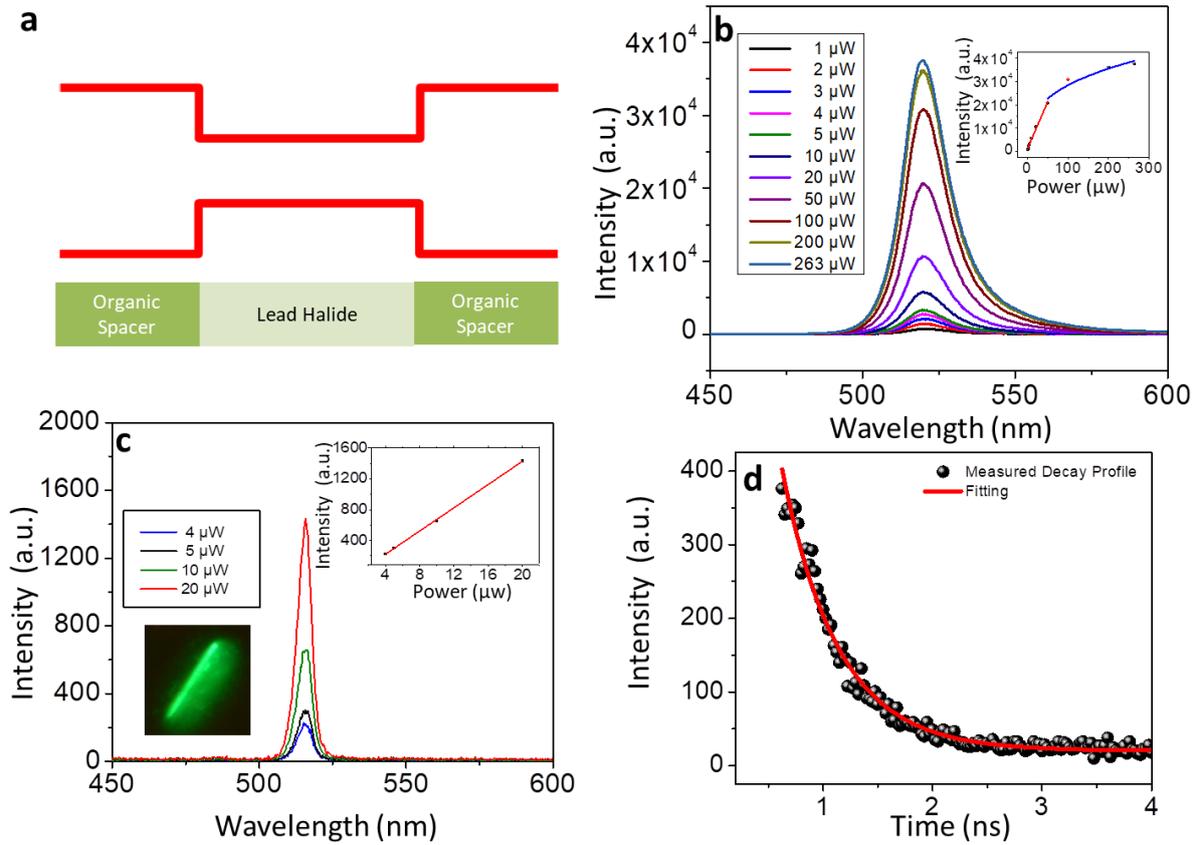

**Figure 3.** Optical properties of the 2D perovskite nanowires (a) Schematic band structure of 2D perovskite showing natural quantum confinement. (b) PL spectra for different excitation power at room temperature. Inset: Integrated PL intensity vs. excitation power. The power dependence has 2 regimes where it shows linear (red fitting) and sub linear (blue fitting) power law dependence with excitation power respectively. The excitation source is a pulsed laser centered at 2.88 eV (430 nm). (c) PL spectra for different excitation power at 77 K. Inset on the left shows the florescence image of the nanowire at 77 K. Inset on right shows a linear power dependence of the integrated PL intensity. The excitation source is a pulsed laser centered at 2.88 eV (430 nm) (d) Time-resolved PL at 77K. The data (black dots) is fitted by a single exponential fit (red) with the lifetime ~ 512 ps.

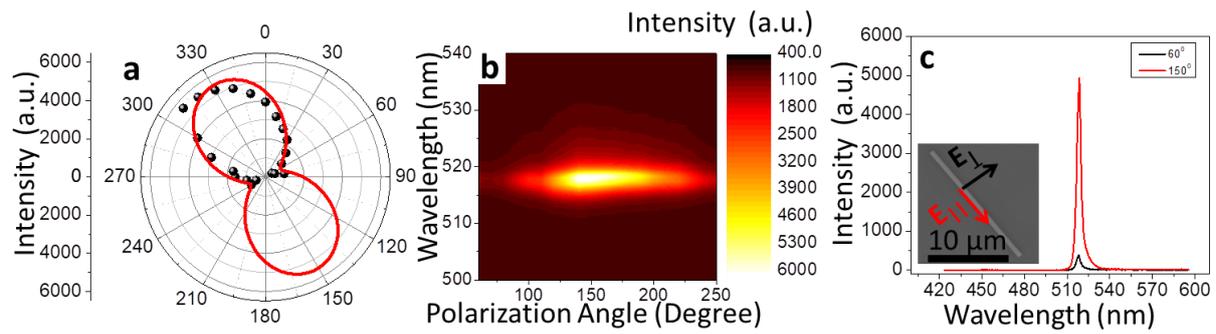

**Figure 4.** Polarization dependence of PL intensity from a single 2D perovskite nanowire. (a) PL intensity as a function of the excitation polarization in a polar plot, and the data (black dots) is fitted with a consine (red) function with period of $2\pi$. (b) PL spectra as a function of the incident light polarization, and the color represents the PL intensity. It is evident that only one sharp PL peak (FWHM of 5 nm) centered at 518 nm is visible. (c) PL spectra with the excitation light polarization parallel (red) and perpendicular (black) with respect to the nanowire orientation.

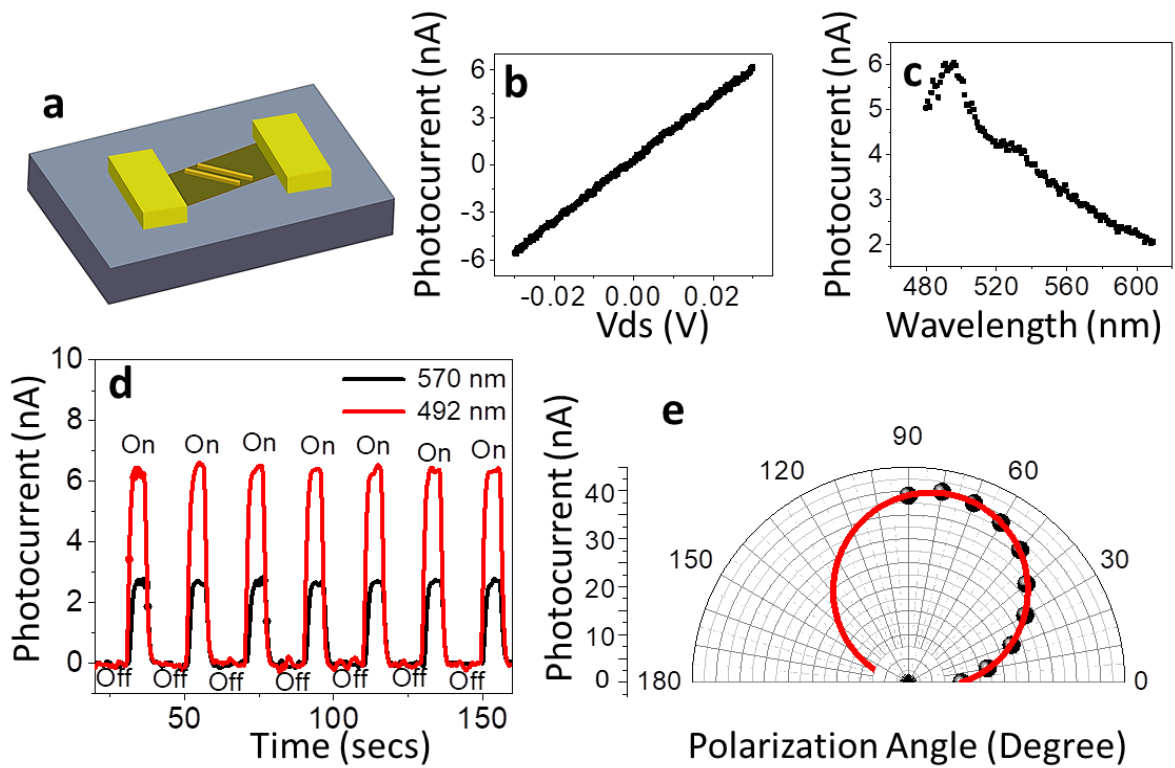

**Figure 5.** Polarization dependence of photocurrent of Graphene/(BA)$_2$PbI$_4$ heterostructure (a) Schematic of the perovskite/grapehene hybride device. (b) Photocurrent response as a function of the source drain voltage (Vds) with pulsed laser centered at 2.52 eV (492 nm) at excitation power of 21 µW. (c) Photocurrent response as a function of the excitation laser energy at a fixed power of 21 µW and Vds of 30 mV. (d) Comparison of normalized photocurrent response when excited with light of different wavelengths with a pulsed laser centered at 2.52 eV (492 nm) at excitation power of 21 µW and bias of 30 mV. (e) Polarization dependence of photocurrent with a pulsed laser centered at 2.52 eV (492 nm) and source drain bias of 30 mV.

Supporting Information

**Catalyst-Free and Morphology-Controlled Growth of 2D Perovskite Nanowires for Polarized Light Detection**

*Debjit Ghoshal[1], Tianmeng Wang[1], Hsin-Zon Tsai[2], Shao-Wen Chang[2], Michael Crommie[2], Nikhil Koratkar[3, 4*] and Su-Fei Shi[1, 5*]*

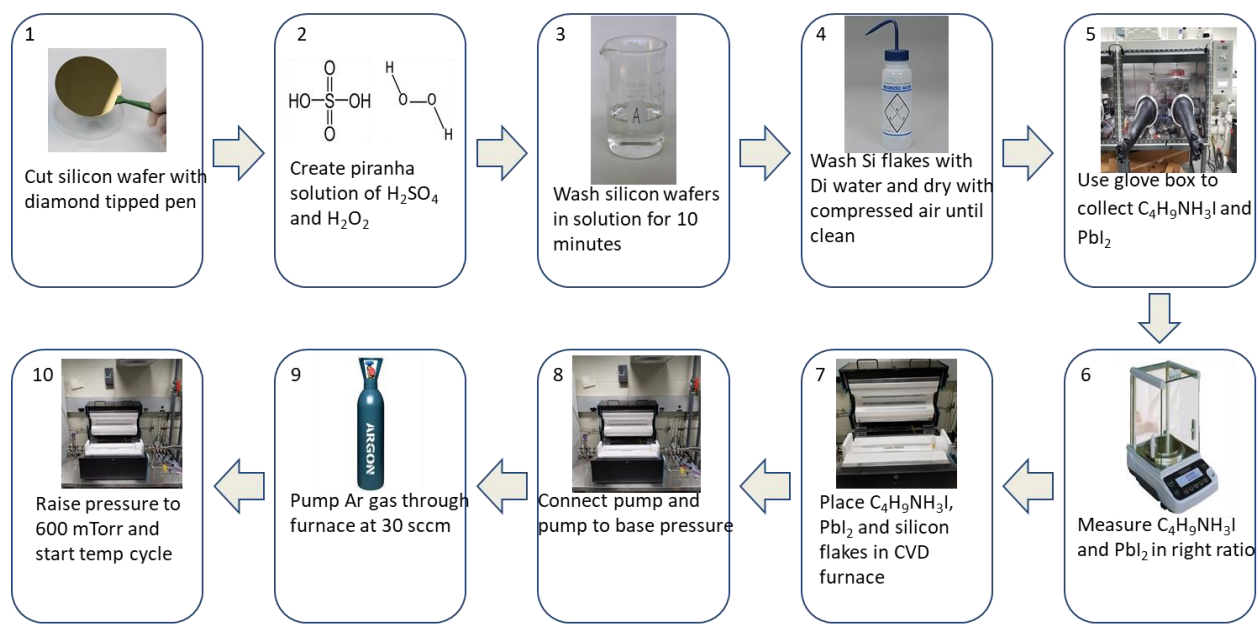

**Figure S1.** Schematic outlining the entire growth process: Steps 1 through 4 involve cleaning the wafer with piranha solution followed by rinsing in acetone and isopropanol. Steps 5-7 involve putting the precursors into an alumina boat for Low Temperature Chemical Vapor Deposition. Steps 8-10 deal with the deposition process which is done at a temperature of $400^0C$ at a pressure of 600 mTorr with Argon as the transport gas which is maintained at a flow rate of 30 sccm during the growth process. The details of the temperature profile in the furnace are listed in Figure S2.

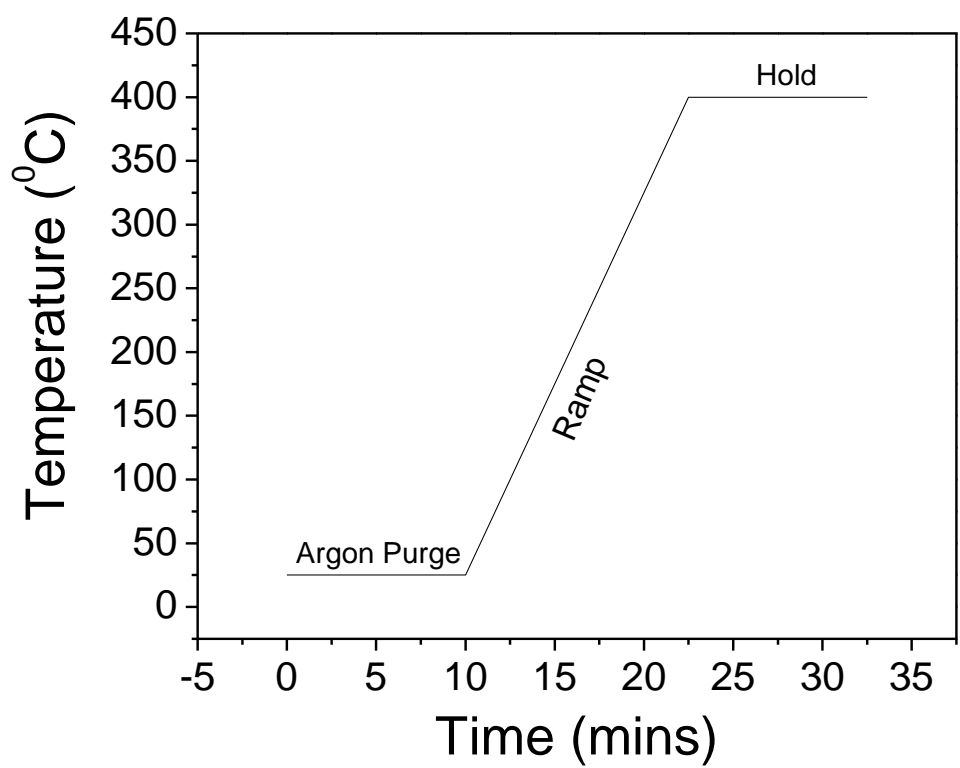

**Figure S2**. Temperature profile in the furnace as a function of time. The furnace is held at a constant temperature (25$^0$C) during the first 10 minutes where it is purged with Argon gas. The second phase is the ramping phase where the furnace ramps up to 400$^0$C at 30$^0$C/min. In the final stage, it is held at 400$^0$C for another 10 mins.

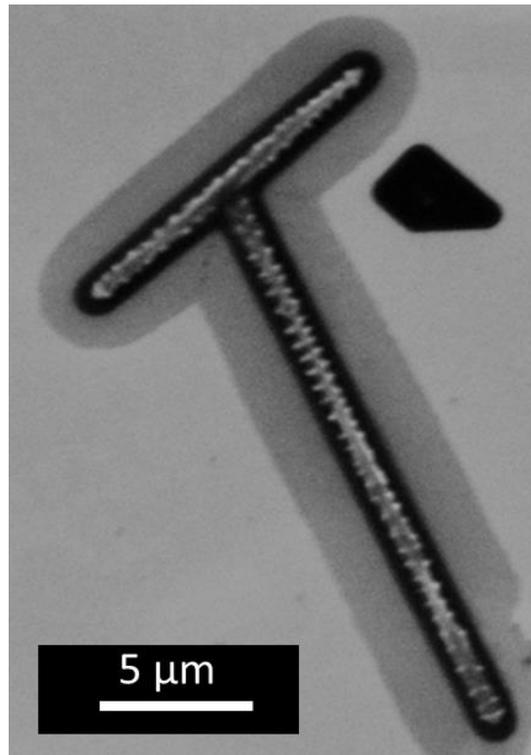

**Figure S3**. SEM images showing spiraling nature of the growth, suggesting Screw Dislocation Driven growth as the most likely mechanism for growth

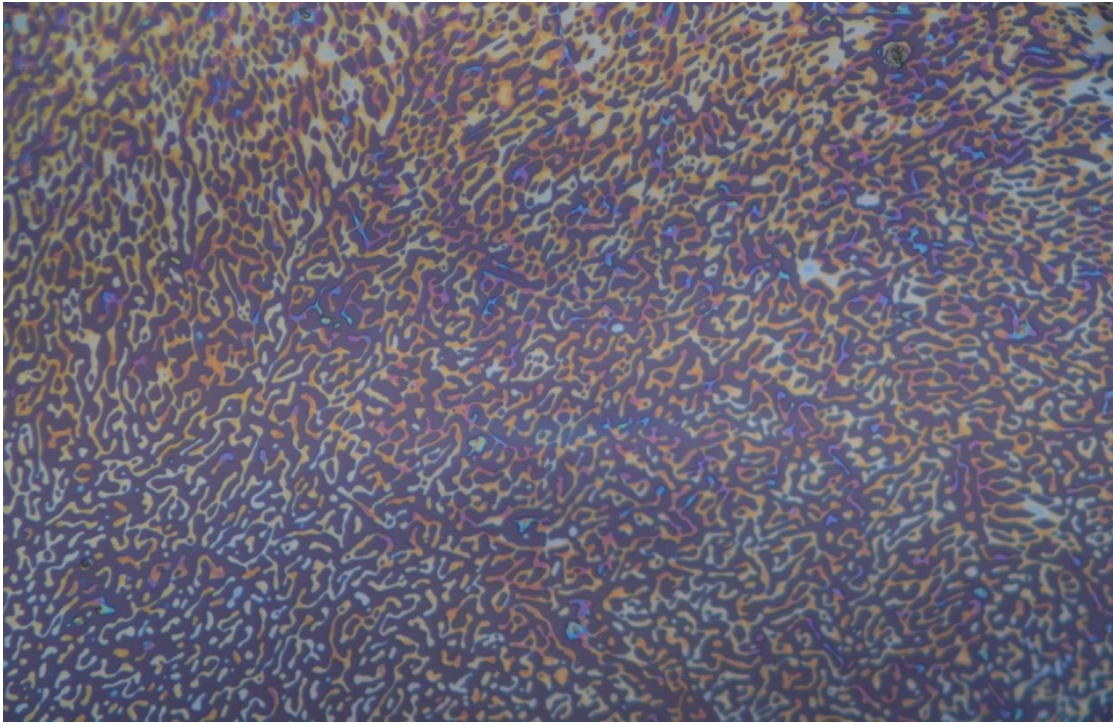
**Figure S4**. Optical Image of dendritic growth at extremely high initial precursor concentrations closest to the furnace.

a

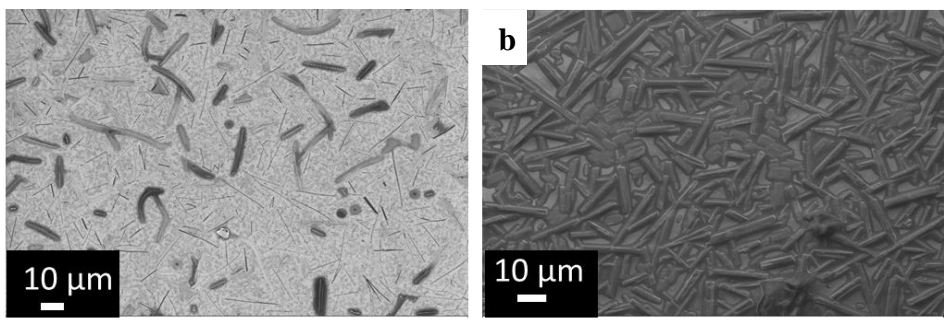

**Figure S5.** SEM Images showing (BA)$_2$PbI$_4$ nanowires of different dimensions grown by LTCVD.

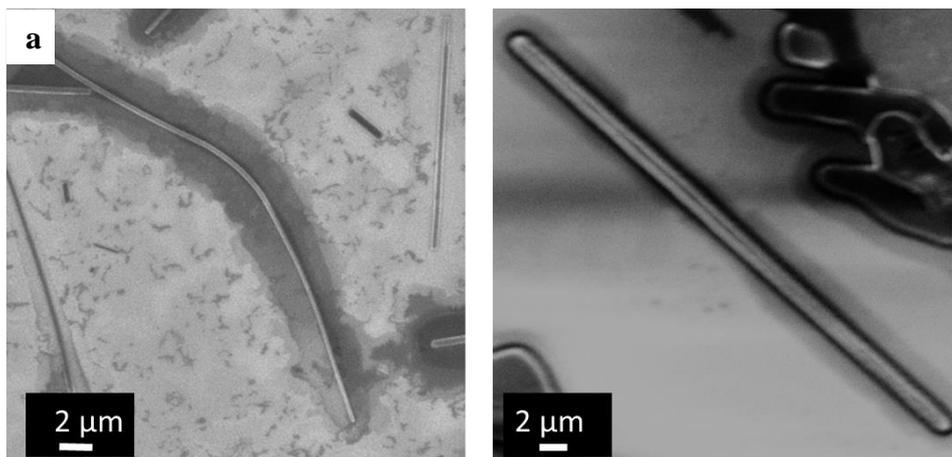

**Figure S6**. Higher magnification SEM images showing (BA)$_2$PbI$_4$ nanowires of different dimensions grown by LTCVD.

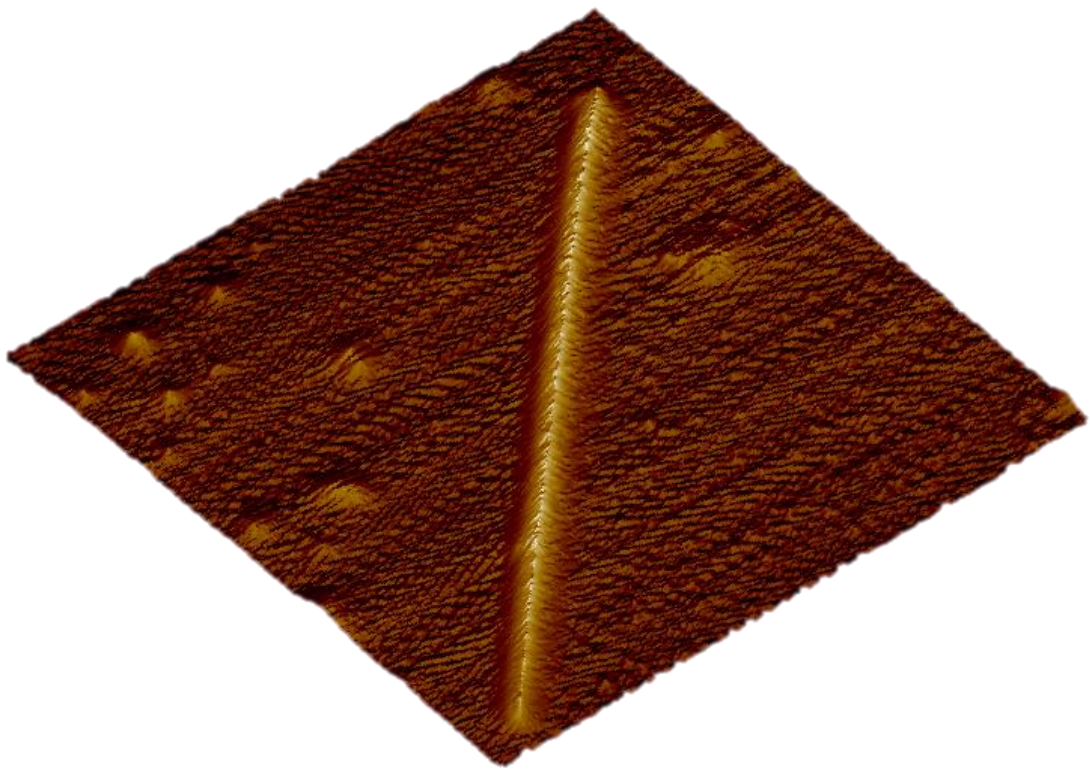

**Figure S7**. Surface AFM 3D image of the nanowire shown in Figure 2a. The scan area is 10 μm x 10 μm. The 3D AFM image shows uniform topography of the nanowire. The measurements were doing using a VEECO MULTIMODE AFM tool in the tapping mode.

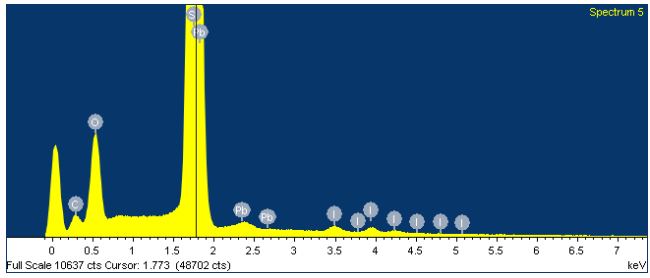

**Figure S8**. EDAX spectrum for the nanowire showing the presence of Pb and I in the right atomic ratio of ~1:4 confirming the identity of the material.

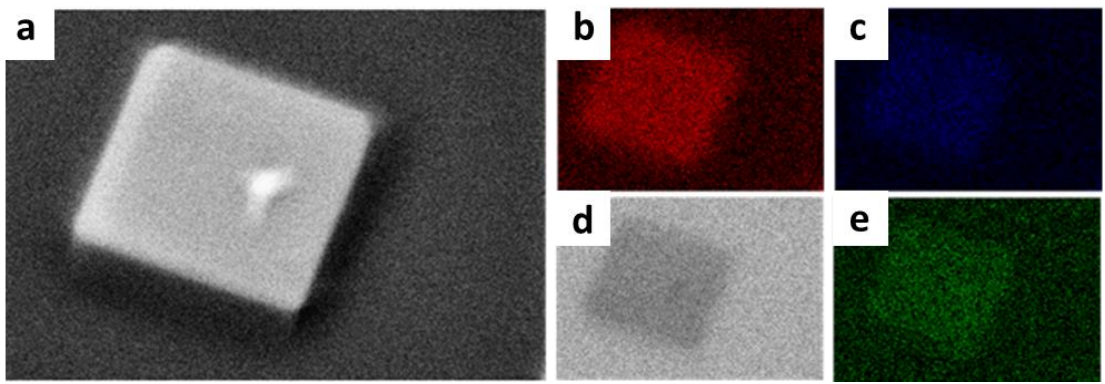

**Figure S9.** Elemental characterization of the nanoplatelets: (a) SEM image of the nanoplatelet (b-e) Elemental mapping of C, I, Si and Pb using EDS spectroscopy. The uniform distribution of the individual elements confirms the homogeneity in chemical composition of the material.

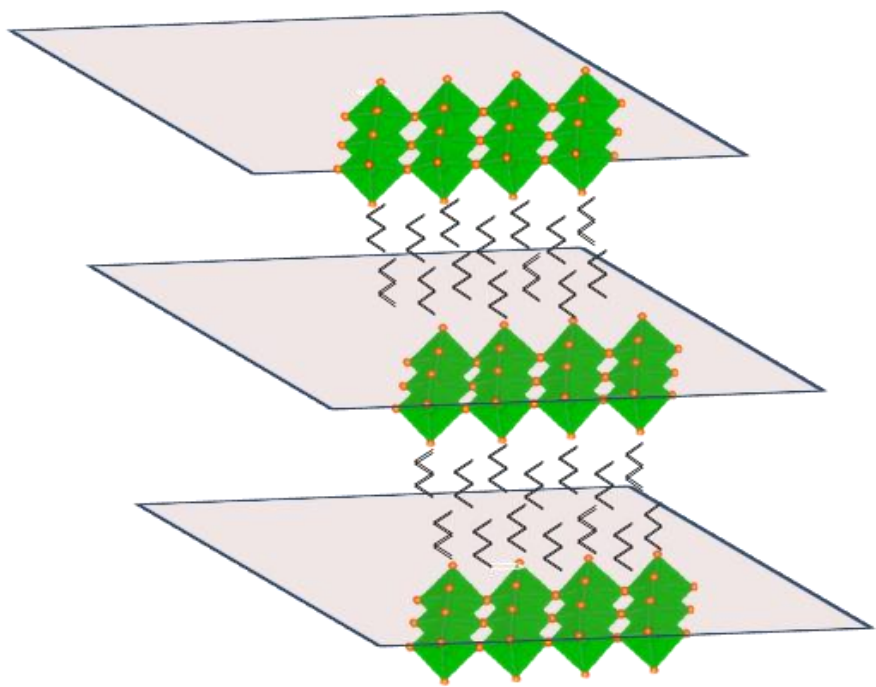

**Figure S10.** Layered Structure of 2D perovskite (BA)$_2$PbI$_4$ with organic spacers separating the Lead Halide layers. The layered structure manifests in a form of a natural quantum confined structure for the material.

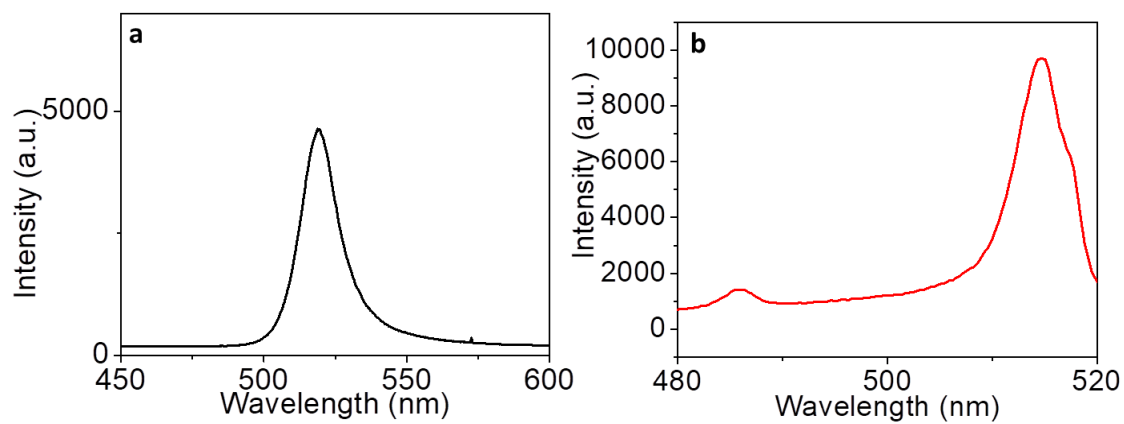

**Figure S11.** PL emission spectrum from nanoplatelet of $(BA)_2PbI_4$. The excitation source is a pulsed laser with a repetition rate of 40 MHz centered at 2.88 eV (430 nm) (a) Room temperature PL showing PL spectrum of nanoplatelets similar to that of nanowires. (b) Low Temperature (77K) PL spectrum showing 2 peaks different from a nanowire.

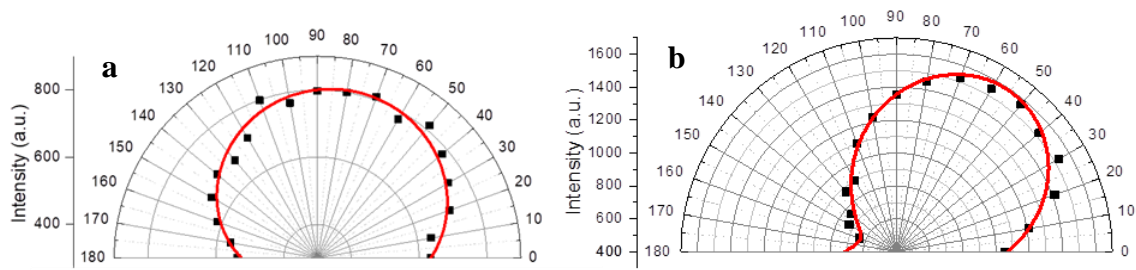

**Figure S12:** Polarization Dependent PL measurements for nanowire samples of different widths (a) 2-3 um (b) hundreds of nanometers

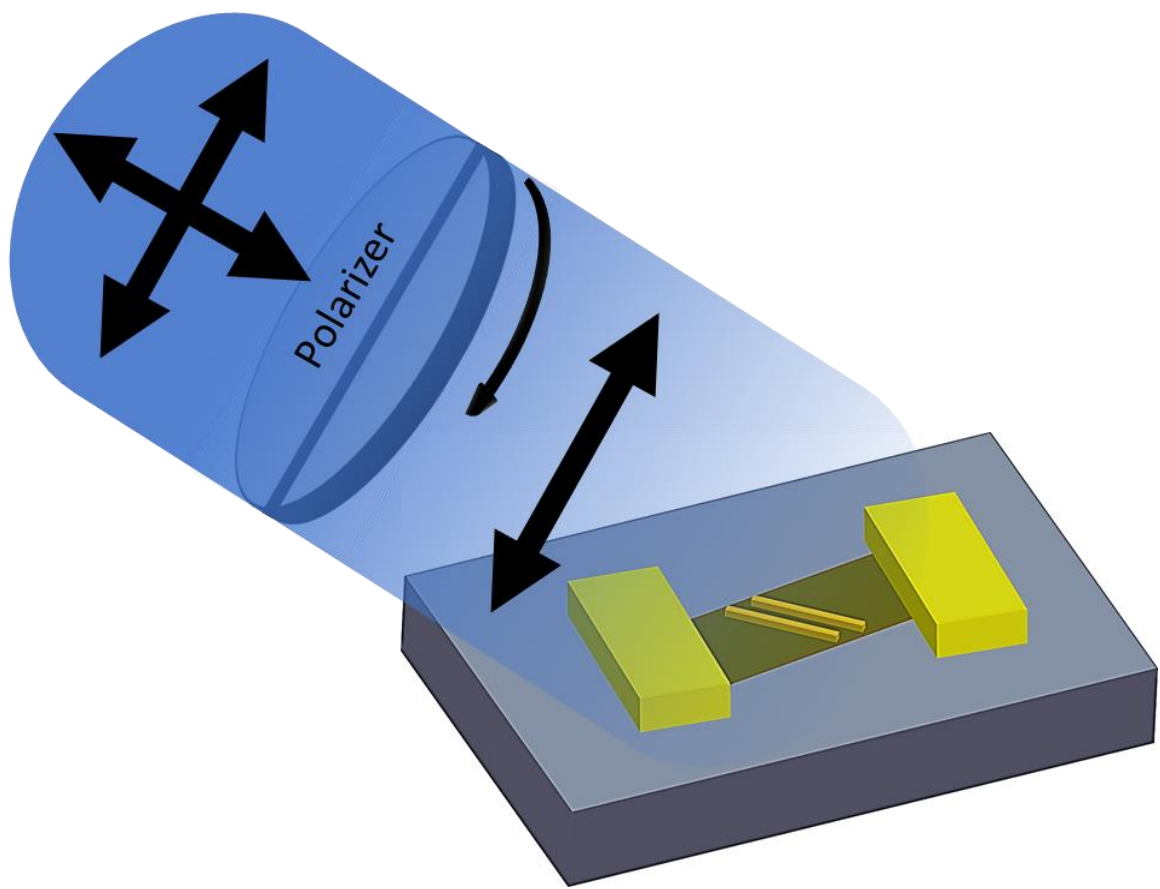

**Figure S13.** Schematic showing the experimental setup for doing polarization dependent photocurrent measurements. The circularly polarized light from a pulsed laser centered at 2.52 eV (492 nm) was passed through a linear polarizer to convert to linearly polarized light. A source drain bias of 30mV is applied.